\documentclass[prb,aps,twocolumn,groupedaddress,floats,showpacs,final]{revtex4}
\usepackage{graphicx}
\usepackage{dcolumn}
\usepackage{bm}
\usepackage{color}
\usepackage{ulem}
\definecolor{blue}{rgb}{0.3,0.3,0.9}
\def\he4{$^4$He}
\def\hee3{$^3$He}
\def\beq{\begin{eqnarray}}
\def\eeq{\end{eqnarray}}

\usepackage[fleqn]{amsmath}
\usepackage{mathrsfs}

\begin{document}

\title{Phases of superclimbing dislocation with long-range interaction between jogs}

\author {Longxiang Liu}
\email{llx1991@mail.ustc.edu.cn}
\affiliation{Department of Modern Physics, University of Science and Technology of China, Hefei, Anhui 230026, China and Department of Engineering \& Physics and the Graduate Center, CUNY, Staten Island, NY 10314, USA}
\author{Anatoly B. Kuklov}
\affiliation {Department of Engineering \& Physics and the Graduate Center, CUNY, Staten Island, NY 10314, USA}

\date{\today}
\begin{abstract}
The main candidate for the superfluid pathways in solid \he4 are dislocations with Burgers vector along the {\it hcp} symmetry axis. Here we focus on quantum behavior of a generic edge dislocation which can perform superclimb -- climb supported by the superflow along its core. The role of the long range elastic interactions between jogs is addressed by Monte Carlo simulations. It is found that such interactions do not change qualitatively the phase diagram found without accounting for such forces. Their main effect consists of  renormalizing the effective scale determining  compressibility of the dislocation in the Tomonaga-Luttinger Liquid phase.  It is also found that the quantum rough phase of the dislocation can be well described within the gaussian approximation
which features off-diagonal long range order in 1D for the superfluid order parameter along the core.  
\end{abstract}

\maketitle

\section{Introduction}
Dislocations are linear topological defects in  crystals which determine the amazing variety of properties of real materials (see in Ref. \cite{Nabarro}). In most cases dislocations are described as classical strings producing long range strain and stress around their cores. This stress is responsible for interactions between dislocations and, correspondingly, for the emerging collective structures and  the strongly non-linear dynamics -- classical plasticity. A complete description of dislocation ensembles remains a tantalizing technological problem which is also of fundamental importance. 

The role of quantum mechanics in dislocation dynamics has also been discussed.  Generating kink-antikink pairs along dislocation by quantum tunneling under stress has been described in Ref.\cite{Pokrovsky}. However, beyond this result the role of quantum mechanics in dislocation induced plasticity in technological materials remains largely an open question. 
In metals edge dislocation may induce superconductivity by strain  by increasing local temperature of the transition within some radius from its core \cite{Nabutovski}.
This model is based on a phenomenological form of the minimal interaction between isotropic strain and scalar superconducting order parameter, and the experimental observation consistent with the proposal has been reported in Ref.\cite{Khlustik}. It is worth mentioning that in this scenario  the dislocation dynamics is not relevant. 

Simulations of screw dislocation along the $C_6$ symmetry axis in solid \he4 have revealed that its core can be
superfluid at low temperature and pressures close to the melting line \cite{screw}. Symmetry of the problem  indicates that the interaction between the strain field and superfluid order parameter must be of second order with respect to the strain \cite{stress}.
A significant difference with the situation in superconductors is that in solid \he4  the same particles form crystalline order (modified by the dislocation topology) and participate in forming algebraic off-diagonal correlations. In this sense a crystal containing such a dislocation represents an example of a supersolid phase of matter. The experimental observation \cite{Hallock} of the supercritical flow through the solid \he4 is consistent with the simulations -- at least at the qualitative level.

The most dramatic effect where quantum mechanics impacts dislocation dynamics has been observed in simulations of the edge dislocation with Burgers vector along the $C_6$ axis \cite{sclimb}. The dislocation dynamics turned out to be strongly intertwined with the superfluidity along the dislocation core which results in the so called {\it superclimb} effect -- the dislocation climb supported by the superflow along the core. This effect is essentially a mechanism of injecting \he4 atoms into the solid from superfluid with the help of the vycor "electrodes" -- in line with the experimental observation of the so called {\it syringe } effect \cite{Hallock_2010}. According to the superclimb mechanism one dislocation climbing across a sample can supply (remove) one layer of atoms. 
 
As discussed in Ref.\cite{sclimb} within the gaussian approximation,  a generic superclimbing dislocation (that is, tilted in the Peierls potential) is a 1D object characterized by excitation spectrum which is parabolic in the momentum along the core. This occurs because of the gauge-type interaction between the superflow and the lateral motion of the core. Thus, such a dislocation represents an example of non-Luttinger liquid \cite{PRB2014}. However, recent analysis \cite{Yarmol} of a generic superclimbing dislocation  beyond the guassian approach has found that 
quantum fluctuations can restore the Tomonaga- Luttinger Liquid (TLL) behavior of the dislocation. This implies that superclimb of the dislocation is suppressed in the limit of zero temperature. In other words, the dislocation transforms from thermally rough to quantum smooth state. Furthermore, the phase diagram of the dislocation in the plane of the crystal shear modulus $G$ and the superfluid stiffness $\rho_s$ along the core features a line of the quantum phase transitions -- between TLL and insulator where the superflow along the core becomes suppressed as well.  

The analysis \cite{Yarmol} was based on the string model of dislocation coupled to the superfluid phase \cite{sclimb} which ignores long range elastic forces between dislocation shape fluctuations.   At this juncture it is important to emphasize the crucial role the long range forces play in quantum glide of a dislocation \cite{EPL}. It was found that arbitrary small long range interaction between kinks of the dislocation aligned with Peierls potential    
suppresses the quantum roughening transition. This transition is essentially the same which occurs in a TLL confined in a lattice with
integer filling. The analogy with the superclimbing dislocation, which also undergoes such a transition \cite{Yarmol}, raises the question if long range forces between jogs should also eliminate the transition and produce insulating phase of the superclimbing dislocation.

In this paper we analyze a superclimbing dislocation with long range forces between jogs. Our main result is that, in a sharp contrast with the gliding dislocation \cite{EPL}, all the phases observed in Ref.\cite{Yarmol} for superclimbing dislocation remain qualitatively unaltered by the long range forces. The role of the long-range forces is reduced to the renormalization of the dislocation compressibility as a function of the shear modulus and strength of the long-range forces --  into a single master curve.

\section{Linearized analysis of the superclimb with Coulomb-type interaction}\label{Linear}

\subsection{Dislocation action}

A superclimbing dislocation with its core along $x-$direction and Burgers vector along $z-$direction  can be modeled as an elastic string of length $L$ which can climb in the $y-$direction along the XY-plane. The climb is supported by superflow along the core \cite{sclimb}. Similarly to Ref.\cite{Yarmol}, we consider dislocation with finite density of jogs of one sign -- that is, a dislocation which is tilted with respect to the Peierls potential rendering this potential essentially irrelevant \cite{PRB2014, Yarmol}. The corresponding action in imaginary time 
\beq
S=S_0[\phi,y] + S_{int}[y],
\label{Stot}
\eeq
is a functional of two variables: $y=y(x,\tau)$ describing position of the dislocation in the XY-plane and imaginary time $\tau$, and   the superfluid phase $\phi=\phi(x,\tau)$ defined along the core.  
Here
\begin{eqnarray}
S_0[\phi,y]&=\int_{0}^{\beta}d\tau\int_{0}^{L}dx\left[-i\left(y+n_0\right)\partial_{\tau}\phi+\frac{\rho_0}{2}\left(\partial_x\phi\right)^2\right.\nonumber\\
&\left.+\frac{\kappa_0}{2}\left(\partial_{\tau}\phi\right)^2+\frac{G_1}{2}\left(\partial_xy\right)^2-\mu y\right],
\label{S0}
\end{eqnarray}
(in units $\hbar=1$, $K_B=1$) 
stands for the short range part of the action considered in Ref.\cite{Yarmol} with $\beta=1/T$, and
\begin{equation}
S_{int}[y]=\frac{G_2}{2}\int_0^\beta d\tau\int_0^Ldx\int_0^Ldx'\frac{\partial_xy \partial_{x'}y}{\left|x-x'\right|+a},
\label{Sint}
\end{equation}
describes the long range interaction between jogs , with $a$ being a short range cutoff (of the order of interatomic distance). This interaction is induced by exchanging bulk phonons between parts of the string separated by a distance $x-x' \,\,$ \cite{Hirth,Kosevich}.
The other notations used in Eqs.(\ref{S0},\ref{Sint}) are as follows: 
$\rho_0, \kappa_0$ are superfluid stiffness and compressibility, respectively; 
$n_0$ stands for the average filling factors; the  parameters $G_{1,2}$ are determined by crystal shear modulus and symmetry (we consider the isotropic approximation); $\mu$ is external bias by chemical potential counted from the value at which the dislocation is in its equilibrium position $y=0$.
The imaginary term in $S_0$, Eq.(\ref{S0}), is the gauge-type interaction between $y$ and $\phi$ leading to the superclimb and non-TLL behavior.

We impose the boundary condition $y(0,\tau)=y(L,\tau)= 0$ in order to avoid the zero mode which corresponds to uniform shift of the string (costing no energy).
 Since we are considering the limit of low (Matsubara) frequency, $\omega\rightarrow 0$, and large wavelengths, $q\rightarrow 0$, we omit the kinetic energy term $\sim(\partial_\tau y)^2$ of the dislocation climb because the main contribution to the kinetic energy comes from the superflow along the dislocation (X-direction) in this limit.

Full statistical description of the dislocation implies evaluation of the partition function 
\beq
Z=\int \mathscr{D}\phi \mathscr{D} y \exp(-S)
\label{Z0}
\eeq 
as the functional integral over $\phi$ and $y$,
where the compact nature of the phase $\phi$ (that is, the possibility of existence of instantons)  must be taken into account. 

\subsection{Gaussian approximation}\label{sec:gauss}
The action (\ref{Stot}) can be analyzed in gaussian approximation by ignoring the compact nature of the phase $\phi$ (and, thus, treating it as a gaussian variable). Then, it is straightforward to obtain spectrum of the excitations from the variational equations of motion $\delta S/\delta y=0,\, \delta S/\delta \phi=0$ : 
\beq
-  i \partial_\tau \phi - G_1 \partial^2_x y - G_2 \partial_x \int dx' \frac{\partial_{x'} y}{|x-x'|+a}= \mu,
\label{dSy}
\eeq
\beq
 i \partial_\tau y - \rho_0  \partial^2_x \phi  - \kappa_0  \partial^2_\tau \phi =0 .
\label{dSphi}
\eeq
Since we are interested in the low energy limit, the last term in Eq.(\ref{dSphi}) can be dropped. Then, we arrive at
\beq
\partial^2_\tau \phi - G_1\rho_0  \partial^4_x \phi - G_2\rho_0 \partial_x \int dx' \frac{\partial^3_{x'} \phi}{|x-x'|+a}=0.
\label{dSpp}
\eeq
As discussed in Refs.\cite{sclimb,Yarmol} for $G_2=0$ this corresponds to the parabolic spectrum $\omega= \sqrt{G_1\rho_0} q^2$ with respect to the momentum $q$ along the core, where $\omega$ corresponds to frequency in real time.  At finite $G_2$ this spectrum acquires the logarithmic correction $\omega = \sqrt{\rho_0 (G_1 +G_2 \gamma \ln(1 + 1/(qa)^2)} q^2$, where the Fourier transform of the long range kernel $1/[|x-x'|+a]$ is taken as $\approx \gamma \ln(1 + 1/(qa)^2)$ with $\gamma \sim 1$.

Eq.(\ref{dSpp}) should be compared with the standard TLL equation of motion
\beq
\kappa_0\partial^2_\tau \phi + \rho_0  \partial^2_x \phi=0
\label{LL}
\eeq
(in imaginary time) in the absence of the Berry term ($ \sim i y \partial_\tau \phi$)  in the action (\ref{S0}). The corresponding spectrum (in real time) $\omega=\sqrt{\rho_0/\kappa_0}q$ is linear in $q$. 

The parabolic spectrum of  superclimbing dislocation following from Eq.(\ref{dSpp}) can be interpreted in terms of the diverging compressibility $\kappa$ -- the {\it giant isochoric compressibility} \cite{sclimb}. In Fourier $ \kappa^{-1} = [G_1 +G_2 \gamma \ln(1 + 1/(qa)^2] q^2$, which leads to the divergence of $\kappa$  for the longest wavelength $q \approx 1/L$   as
\beq
\kappa  \approx \frac{ L^2}{G_1 +G_2 \gamma \ln(1 +( L/a)^2)},
\label{kappa_giant}
\eeq
or $\kappa \sim L^2/[G_2\ln(L/a)]\to \infty$ as $L \to \infty$. 
 
It is important to emphasize that the divergence (\ref{kappa_giant}) does {\it not} imply that a 3D sample permeated by a network of such dislocations should show a diverging 3D compressibility. As discussed in Refs.\cite{PRB2015,Yarmol} for $G_2=0$, the diverging $\kappa$ for one dislocation  means that a sample of solid \he4 permeated by a uniform network of superclimbing dislocations exhibits a linear 3D response on chemical potential which is {\it independent} of the dislocation density, with its magnitude being comparable with the 3D compressibility of a liquid. This property is the basis for the syringe effect \cite{sclimb,Hallock_2010} . 

At finite $G_2$ the 3D response becomes  suppressed  logarithmically with respect to a typical length $L$ of superclimbing segments.
Indeed, a typical element of the network of volume $\sim L^3$ can acquire (or lose) $\sim yL$ extra particles due to the bias $\mu \neq0 $. The value of $y$ in the quasi static limit follows from Eq.(\ref{dSy}) as $y \sim \mu L^2/(G_2 \ln L)$. Thus, the fractional mass change becomes logarithmically suppressed as $\approx y L/L^3 \sim \mu /(G_2\ln L)$ in the limit $L\to \infty$ of low density $L^{-2} \to 0$ of the superclimbing dislocations. [ Here we do not discuss the possibility of screening of the $\ln L$ term in the ensemble of the dislocations].

\subsection{ODLRO of superclimbing dislocation at $T=0$ }
It is interesting to note that, counter intuitively,  in the superclimbing regime the dislocation is characterized by off-diagonal long range order (ODLRO) {\it not} expected in 1D at $T=0$. To demonstrate this, the density matrix $\langle \psi^*(x,\tau) \psi(x',\tau) \rangle$ of the field $\psi =\exp(i\phi)$ can be calculated within the gaussian approximation (\ref{Stot}-\ref{Z0}). Ignoring the log-corrections we find
\beq
\langle \psi^*(x,\tau) \psi(x',\tau) \rangle= \exp\left(-\frac{\sqrt{G_1}}{2\pi a \sqrt{\rho_s}}\right)
\label{ODLRO}
\eeq 
in the limit $|x-x'| \to \infty$,
where the coordinates $x,x'$ are along the core and  $1/a$ stands for the upper cut off of the momentum integration.

The emergence of the ODLRO in 1D is unexpected. As it is clear from above, it is a direct consequence of the parabolic excitation spectrum of the dislocation. As discussed in Ref.\cite{Yarmol} and will be addressed further  below, this spectrum undergoes a transformation into the linear dispersion in the quantum limit giving rise to the TLL phase -- as long as the external bias $\mu$ is below some threshold.
In this phase  the density matrix demonstrates the standard algebraic order  $ \langle \psi^*(x,\tau) \psi(x',\tau) \rangle \sim 1/|x-x'|^c$, with the exponent determined by
the emerging Luttinger parameter $K_{eff}=\sqrt{\rho_0 \kappa_{eff}}$ as $c=1/(2\pi K_{eff})$ . The value of the effective compressibility $\kappa_{eff}$ will be discussed below. However, as shown in Ref.\cite{Yarmol} and will also be discussed below,
the bias $\mu$ can destroy the TLL phase by inducing the quantum rough phase of the dislocation -- that is, the phase characterized by the superclimb. Accordingly, the ODLRO is reinstated at $T=0$.

It should be mentioned that, in contrast to 3D, this ODLRO is fragile -- at any finite temperature $T$ the density matrix becomes exponentially decaying as
\beq
\langle \psi^*(x,\tau) \psi(x',\tau) \rangle= \exp\left(-\frac{T|x-x'|}{2\pi \rho_0}\right),
\label{FT}
\eeq 
in the limit $|x-x'| \geq \sqrt{\sqrt{G_1\rho_0}/T}$. 

As discussed in Ref.\cite{Yarmol}, the linearized analysis of the system does not describe the effect of emergence of the TLL and insulating behaviors as $T\to 0$ and $L \to \infty$. The compact nature of the superfluid phase needs to be taken into account.   
This can be done in the dual representation as explained in the following sections. 

\section{Dual description }\label{dual}

In order to go beyond the gaussian approximation by allowing  instantons, we discretize the space-time into sites $(x,\tau)$ on square lattice, and take into account the compact nature of the phase $\phi$.This implies transforming the integration $\int d\tau \int dx...$ into the summation $\sum_{\tau}\sum_x\Delta\tau\Delta x...$ over the space-time lattice. Specially, we set $\Delta x=a$ and select $a$ as unit of length naturally determined by a typical interatomic distance. The imaginary time increment  $\Delta\tau=\beta/N_\tau$ is determined by the number of time slices $N_\tau$.  Correspondingly, the continuous derivatives $\partial_x\phi(x,\tau)$, $\partial_{\tau}\phi(x,\tau)$ and $\partial_x y$ transform to $\nabla_x\phi(x,\tau)\equiv\phi(x+1,\tau)-\phi(x,\tau)$, $\partial_\tau \phi \to \nabla_\tau\phi(x,\tau)/\Delta \tau$, with $ \nabla_\tau \phi \equiv \phi(x,\tau+\Delta\tau)-\phi(x,\tau)$ and $ \nabla_x y\equiv y(x+1,\tau)-y(x,\tau)$. Then, the action (\ref{Stot}) becomes
\begin{eqnarray}\label{Sdis}
&&S(\phi,y)=\sum_{(x,\tau)} \left[-i(y+n_0)\nabla_\tau\phi+\frac{\Delta \tau\rho_0}{2}(\nabla_x\phi)^2\right.\nonumber\\
&&\left.+\frac{\kappa_0}{2\Delta \tau}(\nabla_\tau\phi)^2+\frac{\Delta \tau G_1}{2}(\nabla_x y)^2\right.\nonumber\\
&&\left.+\frac{\Delta \tau G_2}{2}\sum_{x'}\frac{\nabla_xy\nabla_{x'}y}{|x-x'|+1}-\Delta \tau\mu y\right]
\end{eqnarray}
where the limit  $N_\tau\rightarrow\infty$ at fixed $\beta$ should be approached.

Compactness of $\phi$ can be taken into account within the Villain approximation\cite{Villain} $\vec{\nabla}\phi\rightarrow\vec{\nabla}\phi+2\pi\vec{m}$ with $\vec{m}$ being  integer vector variables defined on bonds between neighboring sites. Then, $\phi$ can be regarded as a non-compact gaussian variable. Thus the action (\ref{Sdis}) can be written as
\begin{eqnarray}
&&S(\phi,y,m_x,m_\tau)=\sum_{(x,\tau)}\left[-i(y+n_0)(\nabla_\tau\phi+2\pi m_\tau)\right.\nonumber\\
&&\left.+\frac{\Delta\tau\rho_0}{2}(\nabla_x\phi+2\pi m_x)^2+\frac{\kappa_0}{2\Delta\tau}(\nabla_\tau\phi+2\pi m_\tau)^2\right. \nonumber\\
&&\left.+\sum_{x'} \frac{\Delta\tau (G_1\delta_{x,x'}+ G_2)}{2}\frac{\nabla_xy\nabla_{x'}y}{\left|x-x'\right|+1}- \Delta\tau\mu y\right].
\label{S}
\end{eqnarray}
And the partition function becomes
\begin{eqnarray}
Z=\sum_{m_x , m_\tau}\int \mathscr{D}y\int\mathscr{D}\phi e^{-S(\phi,y,m_x,m_{\tau})}.
\label{Zf}
\end{eqnarray}

The Poisson identity $\sum_m f(m)\equiv \sum_J\int dmf(m)e^{2\pi imJ}$ allows tracing out all $m_x$ and $m_\tau$ at each bond between neighboring sites and also explicitly integrating out the phase variable. Furthermore, similarly to the approach in Ref.\cite{Yarmol}, we focus on the long-wave limit by retaining only the lowest order of spatial derivatives.   
Then, the partition function (\ref{Zf},\ref{S}) finally becomes
\begin{eqnarray}
Z=\sum_{\{J_x\}}\sum_{\{J_\tau\}}e^{-S_J} 
\label{Zd}
\end{eqnarray}
(up to a constant factor), where $J_x=J_x(x,\tau)$ stands for integer current oriented from the site $(x,\tau)$ along X-bond toward the
site $(x+1,\tau)$ ; similarly, $J_\tau=J_\tau(x,\tau)$ is an integer current along the time bond between the sites $(x,\tau)$
and $(x,\tau + \Delta \tau)$; [ Both $J_x$ and $J_\tau$ can be positive or negative]; 
and 
\begin{eqnarray}
&&S_J=\sum_{(x,\tau)}\left[\frac{1}{2\widetilde{\rho}_0}(J_x)^2-\widetilde{\mu} J_\tau
\right.\nonumber\\
&&\left. + \frac{1}{2}\sum_{x'}\left(\widetilde{G}_1\delta_{x,x'} + \widetilde{G}_2\right)\frac{\nabla_x J_\tau \nabla_{x'}J_\tau}{\left|x-x'\right|+1}\right], 
\label{dual_action}
\end{eqnarray}
where $\widetilde{G}_1=G_1\Delta\tau$, $\widetilde{G}_2=G_2\Delta\tau$, $\widetilde{\mu}=\mu\Delta\tau$ and $\widetilde{\rho}_0=1/\left[2\ln(2/\rho_0\Delta\tau)\right]$ (in the limit $\Delta\tau\rightarrow 0$)\cite{Villain}.

As discussed in Ref.\cite{Yarmol}, the qualitative structure of the results does not change in the limit $\Delta \tau \to 0$.
Thus, in order to understand the main phases of the dislocation it is sufficient to consider $\Delta \tau$ fixed as, say, $\Delta \tau=1$ (in chosen units), and use $\widetilde{G}_1=G_1$, $\widetilde{G}_2=G_2$, $\widetilde{\mu}=\mu$ and $\widetilde{\rho}_0=\rho_0$ in Eq.(\ref{dual_action}).

The integration of the $\phi$-variable results in the local constraint 
which is Kirchhoff's current conservation rule: 
\begin{eqnarray}\label{DJ}
\vec{\nabla}\cdot\vec{J}=0,
\eeq
where the discrete divergence is defined as
$ \vec{\nabla}\cdot\vec{J}= J_x(x+1,\tau)-J_x(x,\tau)+J_\tau (x,\tau+1)-J_\tau (x,\tau)$.
This means that the physical configuration space contributing to $Z$ consists of closed loops of the J-currents -- exactly akin to the J-current model introduced in Ref.\cite{Jcur}. 
We emphasize that the model (\ref{Zd},\ref{dual_action},\ref{DJ}) represents a dual version of the original model (\ref{Z0},\ref{Stot},\ref{S0},\ref{Sint}) -- where the original continuous variables (with the phase $\phi$ being defined {\it modulo} $2\pi$) are replaced by the discrete bond currents $J_x, \, J_\tau$ and the constraint (\ref{DJ}).

\subsection{Linear response}
The linear response  of the system is described in terms of the renormalized superfluid stiffness \cite{E.}
\beq
 \rho_s = \frac{L}{\beta} \langle W_x^2\rangle, 
\label{rhos}
\eeq
and the renormalized compressibility 
\beq
\kappa= - \frac{\beta}{ L} \frac{\partial^2 \ln Z}{\partial \mu^2}=\frac{\beta}{L}[ \langle W^2_\tau \rangle - \langle W_\tau \rangle^2 ] .
\label{kappa}
\eeq
The quantities $W_x=\frac{1}{L} \sum_{(x,\tau)} J_x(x,\tau)$, $W_\tau=N_\tau^{-1}  \sum_{(x,\tau)} J_\tau (x,\tau) $ are integers and have the geometrical meaning of windings of the lines formed by the J-currents. By the construction $W_\tau$ is also the total particle number $N$ in the system.
The windings numbers are topological characteristics of a particular configuration of J-currents.

It is also convenient to introduce the quantity 
\beq
\kappa_1= \frac{\langle N \rangle}{L\mu}= \frac{\langle W_\tau \rangle}{L\mu}.
\label{kappa1}
\eeq
 Both $\kappa$ and $\kappa_1$ coincide with each other as $\mu \to 0$. In general, $\kappa, \kappa_1$ are related by the exact formula $\kappa =d ( \mu  \kappa_1 )/d \mu$. Despite that, statistical errors of simulations can be quite different for both quantities. 

The dual formulation of the model is especially effective for numerical purposes. In what follows we will present results of the simulations performed by the Worm Algorithm \cite{WA}.

\section{Phases of superclimbing dislocation}
The action (\ref{dual_action}) has been analyzed in Ref.\cite{Yarmol} in the absence of the long range term, that is, for the case $\widetilde{G}_2=0$. 
The main result of this study is that, as $L$ and $\beta$ both increase, the non-TLL phase crosses over to either TLL or insulator regardless of the filling factor. The line of Berezinskii-Kosterlitz-Thouless (BKT) transition separates both phases in the plane $(\rho_0, G_1)$ \cite{Yarmol}. 

As discussed in Ref.\cite{Yarmol}, the BKT transition should not occur in this system according to the elementary analysis based on counting of the scaling dimensions.  The "paradox" could be resolved if the discrete nature of the variables $J_x,\,J_\tau$  is taken into account \cite{Borya}: 
as $\rho^{-1}_0$ or $G_1$ increases the discrete gradient term $\sim (\nabla_x J_\tau)^2$ in Eq.(\ref{dual_action}) becomes effectively $ \sim J_\tau^2\,\,$.
This implies the standard XY model behavior corresponding to integer filling. Accordingly, the BKT transition should be expected.  In this context, then, it is worth recalling the result \cite{EPL} where it was shown that the long-range forces suppress quantum roughening of gliding dislocation aligned with Peierls potential. Such a dislocation is formally described by the XY  model (despite that there is no superfluid core), and the suppression of the roughening is interpreted as the insulating state of the effective Luttinger Liquid of kinks. Furthermore, the insulating state of kinks has been shown to emerge at arbitrary small value of the long-range interaction. In other words, the long range interaction eliminates the BKT transition in this system \cite{EPL}.

Thus, the question arises if the same forces in the action (\ref{dual_action}) should suppress the superfluidity along the core of the superclimbing dislocation -- also at arbitrary small value of $G_2$.      Clearly, if $\nabla_x J_\tau$ is replaced by $\sim J_\tau$ in the action (\ref{dual_action}) one would arrive at, practically, the same action studied in Ref. \cite{EPL}. Then, the answer would be positive to the above question. 

 However, our numerical results for the model (\ref{dual_action}) contradict to this logic. More specifically, we find that there is a separatrix in the finite scaling behavior which occurs at finite value of $G_2$ of the order of unity. This separatrix indicates the boundary between TLL and the insulator. Furthermore, we show that the effect of finite $G_2$ in Eq.(\ref{dual_action}) is reduced to  renormalization of $G_1$, so that the phase diagram constructed in Ref.\cite{Yarmol} for the case $G_2=0$ can be simply redrawn in terms of the renormalized $G_1$.

\subsection{Renormalized compressibility in the quantum limit }
The compressibilities (\ref{kappa},\ref{kappa1}) show "giant" values $\sim L^2$ at finite $\beta$ as $L \to \infty$ \cite{Yarmol}.
This feature is intimately connected with  the superclimb effect  and the parabolic excitation spectrum \cite{sclimb}. However, simulations of the full model in the limit $\beta \sim L \to \infty$ for $G_2=0$  have found that the compressibility becomes finite if  $G_1$ does not exceed some critical value $G_c$ for a given $\rho_0$. If $G_1>G_c$, the compressibility vanishes which is signaling the insulating behavior.   

\begin{figure}[htbp]
	\centering\includegraphics[width=1.1\columnwidth]{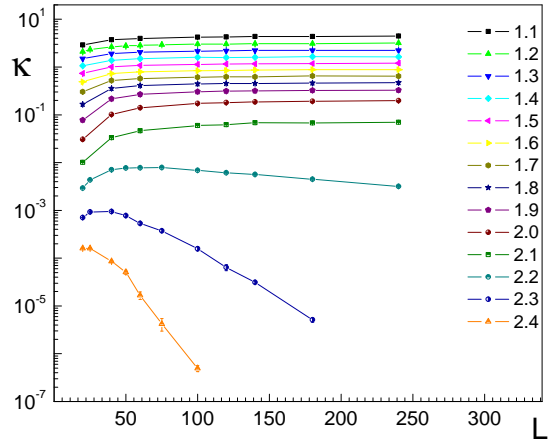}
	\caption{Compressibility $\kappa$ vs $L=1/T$ for various values of $G_1$ shown in the legend at $G_2=1.0$ and $\rho_0=4$, $\mu=0$.}\label{fig:1}
\end{figure}
\begin{figure}[htbp]
	\centering\includegraphics[width=1.1\columnwidth]{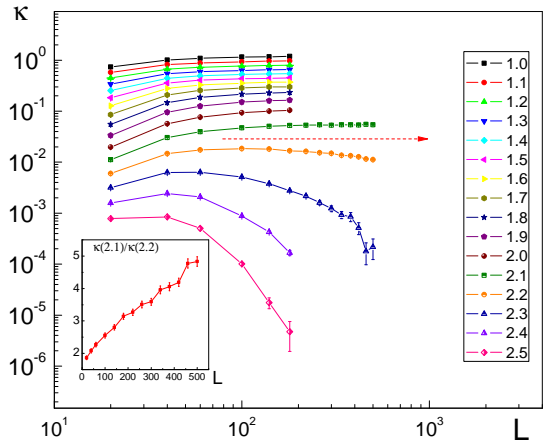}
	\caption{Compressibility $\kappa$ vs $L=1/T$ for various values of $G_2$ shown in the legend at $G_1=1.5$ and $\rho_0=4$, $\mu=0$.The dashed line indicates the approximate position of the separatrix. Insert: the ratio $\kappa(G_1=1.5,G_2=2.1)/\kappa(G_1=1.5,G_2=2.2)$ indicating different types of behavior above and below the separatrix.}\label{fig:3}
\end{figure}
\begin{figure}[htbp]
	\centering\includegraphics[width=1.1\columnwidth]{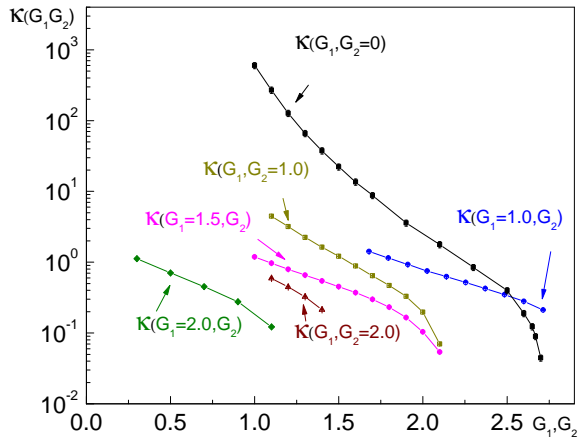}
	\caption{The asymptotic values $\kappa_{eff}$ of $\kappa$ for various values of $G_1$ and $G_2=1.0$. The data for $G_2=0$ are taken from Ref.\cite{Yarmol}.}\label{fig:2}
\end{figure}
\begin{figure}[htbp]
	\centering\includegraphics[width=1.1\columnwidth]{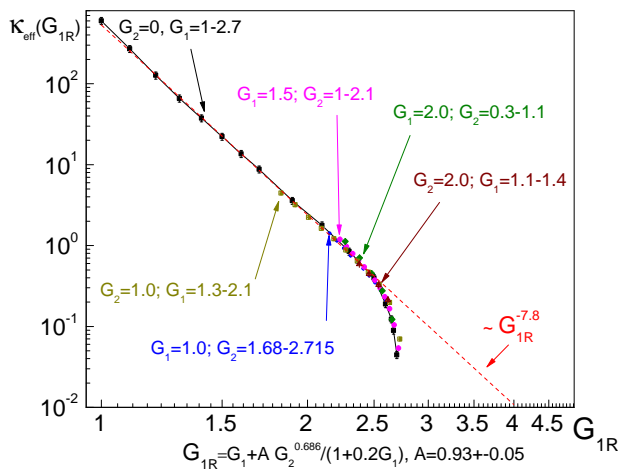}
	\caption{The master curve $\kappa_{eff}$ taken from Fig.~\ref{fig:2} and replotted versus $G_{1R}$. The parameter $A$ has been adjusted for some sets to better fit the master curve within 5\% of deviations, while other parameters were kept fixed for all sets.}\label{fig:master}
\end{figure}

The results of MC simulations performed for finite $G_2$ are shown in Fig. \ref{fig:1}.
It depicts compressibility $\kappa$ at various $L$, with $\beta=L$, and various values of $G_1$ when $G_2=1.0$, $\rho_0=4$ and $\mu=0$. As $L$ increases, $\kappa$ asymptotically approaches some finite value $\kappa_{eff}$, if $G_1$ is below some critical value which can be estimated as $ G_c \approx 2.1$. This behavior is qualitatively the same as observed in Ref.\cite{Yarmol} for $G_2=0$. If $ G_1$ exceeds $G_c$, the compressibility flows to zero  as can be clearly seen in Fig.~\ref{fig:1}. This feature, indicating the quantum transition toward the insulator, is also qualitatively the same as observed in Ref.\cite{Yarmol} for $G_2=0$. Here we didn't study in detail if the transition remains in the BKT universality. Instead, we will give a strong argument in favor the BKT universality at finite $G_2$.
 
The behavior of $\kappa$ vs $L$ for $G_1=1.5$ and varying $G_2$ is shown in Fig.~\ref{fig:3}. The plots also show the saturation to finite values $\kappa=\kappa_{eff}$, if $G_2$ is below some critical value, $G_{2c}\approx 2.1$, and the flow toward the insulator at $G_2>G_{2c}$. In order to emphasize the separatrix type feature (marked by the dashed line), that is, separating the TLL and the insulating  phases, the ratio of $\kappa(G_2=2.1)$, which is showing no visible dependence on $L$ over the extended range, to $\kappa(G_2=2.2)$, which shows deviations from the asymptotic saturation, is presented in the inset to Fig.~\ref{fig:3}.  
A strong divergence of the ratio with growing $L$ emphasizes the  presence of the separatrix.  

The asymptotic values $\kappa_{eff}$ vs $G_1, G_2$ are presented in Fig.~\ref{fig:2} for various combinations of the arguments. 
[The "asymptotic" values of $\kappa$ from the curves Figs.~\ref{fig:1},\ref{fig:3} showing no asymptotic behavior were read off from the largest size simulated]. These curves appear to be unrelated to each other. However, it is important to note that all the data from Fig.~\ref{fig:2} can be collapsed on a single master curve $\kappa_{eff}$ versus the single variable 
\beq
G_{1R}(G_1,G_2)=G_1 + A\frac{G_2^{0.686}}{1+0.2G_1},
\label{GR}
\eeq
 where $A=0.93 \pm 0.05$, which can be viewed as  $G_1$ renormalized in the presence of the long-range interactions. This interpretation is justified because all the data at finite $G_2$ can be collapsed to the curve $\kappa$ vs $G_1$ at $G_2=0$.   The resulting dependence   is shown in Fig.~\ref{fig:master}.   Thus, we conclude that, as long as, $G_{1R}$ is below its critical value $G_c$ (which is $G_c\approx 2.7$ for $\rho_0=4$ ) there is a finite domain of $G_2$ within which the TLL behavior persists. This domain corresponds to the dotted  line $\sim G_{1R}^{-7.8}$ in Fig.~\ref{fig:master}, with the deviations indicating the flow toward the insulating phase.    
Thus, the long range interactions do not change qualitatively the nature of the phase diagram found in Ref.\cite{Yarmol}. Its main role is in renormalizing the $G_1$ value.

\section{Impact of long range forces on superclimb induced by the bias $\mu \neq 0$}\label{rough}

The emergence of TLL behaviour and the corresponding suppression of the superclimb can be viewed from a different perspective. 
The giant compressibility \cite{sclimb,Yarmol} of superclimbing dislocation becomes possible because the dislocation can climb  -- thanks to the supercurrents along the core supplying matter needed to support this non-conservative motion of the core. This determines the rough phase of the dislocation -- when the mean square displacement of the core position exhibits fluctuations logarithmically diverging as $L\to \infty$.
As shown in Ref.\cite{Yarmol} and discussed above, at zero bias by chemical potential, $\mu$, such fluctuations become suppressed in the quantum limit so that the TLL behavior emerges. In other words, the rough phase of the superclimbing dislocation at zero bias can only exist at finite temperature.

The situation is different at finite bias -- the rough phase can be induced by finite $\mu$ in the quantum limit.
This was demonstrated in Ref.\cite{Yarmol} in the case of short range interactions (that is, $\tilde{G}_2=0$ in Eq.(\ref{dual_action})). Furthermore, the dislocation compressibility in this case can be described within the guassian approach treating the dislocation as an elastic string. 
Here we address the question how the bias affects the dislocation in the presence of long range forces.

\begin{figure}[htbp]
	\centering
	\includegraphics[width=1.1\columnwidth]{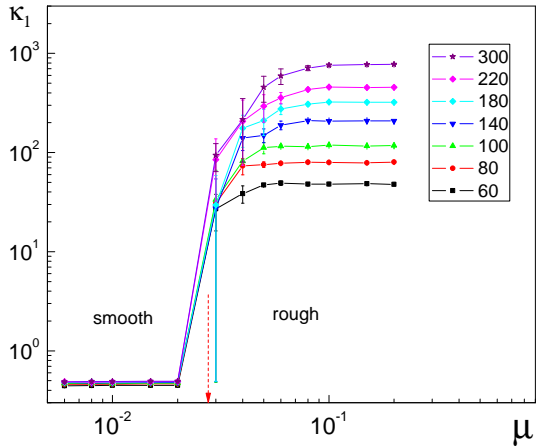}
	\caption{$\kappa_1$ vs $\mu$ for various $L$ up to $L=300$ with $G_1=1.5$, $G_2=1.5$, $\rho_0=4$, $T=0.05$.
	}\label{fig:kappa1_mu}
\end{figure}
\begin{figure}[htbp]
	\centering
	\includegraphics[width=1.0\columnwidth]{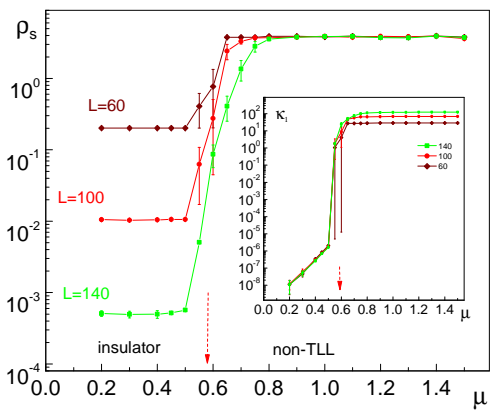}
	\caption{ Superfluid stiffness along the dislocation versus $\mu$ undergoing the transformation from the insulating to the non-TLL phase for different lengths $L$ (shown close to each curve); $T=0.05$.Inset: corresponding $\kappa_1$ versus $\mu$.}\label{fig:rhos}
\end{figure}

 \begin{figure}[htbp]
	\centering
	\includegraphics[width=1.0\columnwidth]{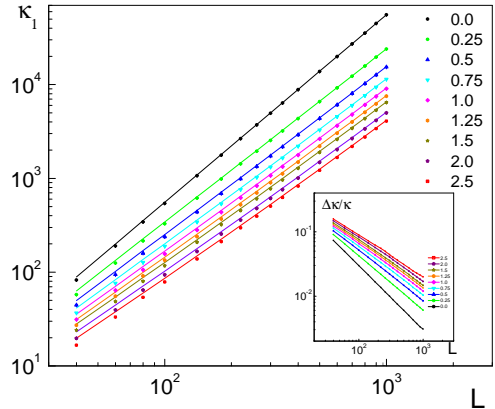}
	\caption{MC data (points) for $\kappa_1$ in the rough state versus the dislocation length  $L$ and for various $G_2$ values with $G_1=1.5$, $1/T=20$ and $\rho_0=4$; The  lines show corresponding results for $\kappa_1$, Eq. \ref{kappa1}, derived within the gaussian approximation (\ref{kappa12}). Inset: the relative deviations between the MC data and the approximation. The decay  is characterized by $\sim L^{-c}$ with some exponent $\sim 1$ ( $c=0.77(2)$ for $G_2=1.00$). }\label{fig:gaussian}
\end{figure}
The results of simulations of the model (\ref{Zd},\ref{dual_action}) at finite $\mu$ and $\tilde{G}_2$  are presented in  Figs.~ \ref{fig:kappa1_mu},\ref{fig:rhos}. As can be seen, the bias induces roughening of the dislocations by restoring the giant compressibility above certain threshold. More specifically, at low values of $\mu$ the dislocation is characterized by  $\kappa$  independent of the dislocation length. This state is marked as "smooth" (which is also the TLL phase)  in Fig.~\ref{fig:kappa1_mu}.  Upon increasing $\mu $ the system undergoes the transformation into the rough phase marked as "rough" (which is also the non-TLL phase) in Fig.~\ref{fig:kappa1_mu} characterized by the value of $\kappa= \kappa_1$ diverging as $L\to \infty$. The curves shown in Fig.~\ref{fig:kappa1_mu} correspond to values of $G_1,G_2$ where the transformation between TLL ("smooth") and non-TLL ("rough") phases takes place. In this case, while $\kappa,\kappa_1$ show dramatic change,  the superfluid stiffness $\rho_s$ remains, practically, unaffected. 
Results of the simulations at $G_{1R}>G_c$,Eq.(\ref{GR}), that is, when the dislocation is in the insulating regime at low $\mu$, are shown in Fig.~\ref{fig:rhos}: As $\mu$ increases, both $\rho_s$ and $\kappa_1$ undergo a strong transformation -- from, practically, zero values at $\mu=0$ (marked as "insulator") to finite $\rho_s$ and giant $\kappa_1$ (marked as "non-TLL") at $\mu$ above the threshold.

\subsection{Compressibility at finite bias in the $T=0$ limit. }

Here we focus on the nature of the quantum rough (that is, the non-TLL)  phase of the dislocation, and will show that this phase can be described quite accurately  within the  gaussian approximation.
In other words, external bias can restore superclimb in the quantum limit. 

Here we compare the results of MC simulations of the full quantum action in the limit where $\kappa$ and $\kappa_1$ show saturation at large $\mu$  (that is, corresponding to  the region $\mu >0.1$ in the graph Fig.~\ref{fig:kappa1_mu}) with the guassian approximation for $\kappa$, Eq.(\ref{kappa}), which can be expressed as
\begin{eqnarray}
\kappa=  \frac{1}{L} \sum_{x,x'} [\langle Y(x) Y(x')\rangle -  \langle Y(x)\rangle \langle Y(x')\rangle] ,
\label{kappa_ga}
\end{eqnarray}
where $Y(x)= (\Delta \tau/\beta) \sum_\tau y(x,\tau)$ corresponds to the Matsubara frequency $\omega=0$. Similarly, using the definition (\ref{kappa1}) one can represent   
 \begin{eqnarray}
\kappa_1= \frac{1}{L \mu} \sum_x \langle Y(x) \rangle.
\label{kappa1_ga}
\end{eqnarray}
The variable $Y(x)$ corresponds to $\omega=0$, and it separates from  higher Matsubara harmonics $\omega$. This allows  evaluating the averages $\langle .... \rangle$ in Eqs.(\ref{kappa_ga},\ref{kappa1_ga}) within the "shortened" action  (\ref{Sdis}) where only three last terms and the harmonic $\omega=0$ are retained. This action, then, takes the form 
\begin{eqnarray}
S_{cl} =&\frac{1}{T} \sum_x\left[ \frac{G_1}{2} (\nabla_x Y_x)^2 - \mu Y_x\right.\nonumber\\
&\left.+\sum_{x'} \frac{G_2}{2(1+|x-x'|)} \nabla_x Y_x \nabla_{x'}Y_x \right],
\label{E/T}
\end{eqnarray}
which is the action for classical string $S_{cl}=E/T$ determined by the potential energy $E$ of elastic deformations.
Accordingly, the statistical averaging is to be performed with the classical gaussian partition function $Z_{cl} = \int DY \exp(-S_{cl})$.

Representing 
\begin{equation}
Y(x)=\sqrt{\frac{2}{L}} \sum^{L-1}_{n=1} \sin(q_n x) f_n 
\label{fn}
\end{equation}
in terms of the spatial harmonics obeying zero boundary condition, where $f_n$ are real variables with $ q_n =\pi n/L,\,\, n=1,2,...,L-1$, and substituting it into Eq.(\ref{E/T}), we find
\begin{eqnarray}
Z_{cl}&=&\int Df_n \exp(-S_{cl}),
\\
S_{cl}&=& \frac{1}{T}\left[\frac{1}{2}\sum_{n,n'} V_{n,n'} f_n f_{n'} - \mu \sum_n \Phi_n f_n \right],\\
\Phi_n&\equiv&  \sqrt{\frac{2}{L}}\frac{(1-(-1)^n)}{2}\cot(\pi n/(2L)),
\end{eqnarray}
where
\begin{eqnarray}
&&V_{n,n'}=G_1(Q_n)^2 \delta_{n,n'} + \frac{2}{L}\sum_{x,x'}   \frac{G_2Q_n Q_{n'}}{1+|x-x'|}\nonumber\\
&\cdot&\cos(q_n(x+1/2))\cos(q_{n'}(x'+1/2)),
\label{V}
\end{eqnarray}
$Q_n\equiv 2\sin(q_n/2)$ and the summations run over $x, x' =0,1, ..., L$. 

The averages (\ref{kappa_ga},\ref{kappa1_ga}) can be expressed as
\begin{equation}
\kappa=\kappa_1= \frac{1}{L} \sum_n \langle \Phi_n f_n \rangle =\frac{1}{L}\sum_{n,n'} \Phi_n (V^{-1})_{n,n'} \Phi_{n'}, 
\label{kappa12}
\end{equation}
where $(V^{-1})_{n,n'}$ is the matrix inverse to $V_{n,n'}$ (which was evaluated by exact diagonalization) \cite{note}.
These values are the compressibilities obtained within the gaussian approximation.

The comparison between this approximation (lines) and the MC data (symbols) are shown in Fig.~\ref{fig:gaussian}. 
 As can be seen, the quality of the gaussian approximation improves as dislocation length increases. 
Thus, it is fare to conclude that the quantum rough phase induced by the bias can  be well described within gaussian approximation, with the deviations reduced below 1\% for sizes  $L> 200-300$ .

\section{Discussion.}
Here we have focused on the stability of the phase diagram of edge dislocation with superfluid core with respect to elastic long-range interactions between jogs. As shown in Ref.\cite{Yarmol} for the case of short-range interactions, such a diagram features  three quantum phases in the space of three parameters $(\rho_0, G_1, \mu)$: i) TLL  which is also the smooth superfluid phase ; ii) the insulator, that is, smooth and non-superfluid; iii) quantum rough -- superclimbing  phase induced by finite bias $\mu$.  As shown above the long-range interactions do not change this picture qualitatively.
The question is why there is such a significant difference between superclimbing and gliding dislocations -- where the long-range interaction eliminates quantum phase transition \cite{EPL}.

It has been shown in Ref.\cite{EPL} that the elastic long-range forces suppress quantum roughening  transition for gliding dislocation
aligned with Peierls potential. In terms of the dual representation of this dislocation by the Coulomb gas approach this means
that the effective interaction between instanton and anti-instanton becomes modified -- from log  to the
log log     of the distance between the instanton pair. This implies that such pairs proliferate at arbitrary small value of the "Coulomb" interaction. Accordingly, the plasma phase of the pairs guarantees that the dislocation is quantum smooth. In other words, arbitrary weak Coulomb-type interaction eliminates the BKT quantum roughening phase transition for gliding dislocation.

Our current numerical results show that the presence of the superfluid core in edge dislocation changes the situation -- the phase diagram of the dislocation retains its structure. At formal level, the difference between two models is easier to understand in terms of the dual representation by the J-currents. In the case of the gliding dislocation \cite{EPL} the duality transformation generates terms with the $\sim 1/r$ interaction between the J-currents. In this sense the "Coulomb" interaction suppresses Luttinger parameter logarithmically and, thus, eliminates the BKT transition for the gliding dislocation for arbitrary small $G_2$.     In contrast, the edge dislocation with superfluid core is described by the model  (\ref{Zd},\ref{dual_action}) where 
the Coulomb-type term acts between spatial derivatives of the J-currents (oriented along imaginary time). Thus this interaction vanishes in the long-wave limit and, accordingly, no suppression of the Luttinger parameter occurs, at least, in the limit $G_2 \to 0$.   
As was discussed above, the role of the long-range forces is reduced to the renormalization of the parameter $G_1$.

An unexpected property of the quantum rough phase is the ODLRO in 1D (along the core). This phase can be induced by the bias $\mu$, and its description can be well achieved within the gaussian model. The exact nature of the transition between TLL (or insulator) and the rough phase is not fully understood. As demonstrated in Ref.\cite{Yarmol}, the transition is characterized by strong hysteresis at low $T$. This indicates Ist order transition which should occur in the limit $T\to 0$. The question is if the transition remains at finite $T$.  In Ref.\cite{Darya} the transition has been analyzed for the dislocation aligned with the Peierls potential,
and the argument has been given that the transition remains at finite $T$ -- in spite of the "no-go" theorem \cite{Landau_V} for a phase transition in 1D at finite $T$. The main argument is that the rough phase is not characterized by any local order parameter with respect to the dislocation shape. Instead, it is a global property of the system which immediately undermines the basis for the theorem \cite{Landau_V}.  Thus, we conjecture that the same argument holds for generic dislocation, and the transition remains at finite $T$.

\noindent {\it Acknowledgments}.
 This work was supported by the National Science Foundation under the grant DMR1720251 and by the China Scholarship Council.


\begin{thebibliography}{99}
\bibitem{note}
The "Coulomb" term ($\sim G_2$) of the interaction matrix (\ref{V})  does not satisfy periodic boundary conditions and, therefore, it cannot be
diagonalized by Fourier transformation.
\end{thebibliography}
\end{document}